# Comparison of Simulation-Guided Design to Closed-Form Power Calculations in Planning a Cluster Randomized Trial with Covariate-Constrained Randomization: A Case Study of Mass Small-Quantity Lipid Nutrient Supplementation Combined with an Expanded Program on Immunization in Rural Chad

**Authors:**

Jay JH Park,[a,b] Rebecca K. Metcalfe,[b] Nathaniel Dyrkton,[b] Yichen Yan,[c] Shomoita Alam,[b] Kevin Phelan,[d] Ibrahim Sana,[d] and Susan Shepherd[d]

**Affiliations:**
  a. Department of Health Research Methods, Evidence, and Impact, McMaster University, 1280 Main Street West, Hamilton, ON, L8S 4L8 Canada
  b. Core Clinical Sciences, 401-34 W. 7th Ave., Vancouver, BC, V5Y 1L6 Canada
  c. Department of Statistical and Actuarial Science, Simon Fraser University, Burnaby, BC, Canada
  d. The Alliance for International Medical Action (ALIMA), Dakar, Senegal.

**Corresponding author:**

Jay JH Park

Department of Health Research Methods, Evidence and Impact

McMaster University

1280 Main St W, Hamilton, ON L8S 4L8

# 1 Abstract

**Introduction:** Current practices for designing cluster-randomized trials (cRCTs) typically rely on closed-form formulas for power calculations. For cRCTs using covariate-constrained randomization, the utility of conventional calculations might be limited, especially for complex cRCTs with multi-nested data structure where the unit of randomization is different from the unit of outcome collection. We compared simulation-based planning of a multi-nested cRCT using covariate-constrained randomization to conventional power calculations using OptiMAx-Chad as a case study.

**Methods:** OptiMAx-Chad is a two-arm, multi-nested cRCT examining the impact of embedding mass distribution of small-quantity lipid-based nutrient supplements within an expanded programme on immunization to improve childhood vaccine coverage in rural villages in Ngouri, Chad. The primary endpoint is the proportion of children aged 12-24 months with at least one dose of measles-containing vaccine (MCV1), measured at the village-level. Within the 12 health areas available for randomization, a random subset of villages will be selected for outcome collection. Using baseline census survey data, 1,000,000 random assignments of health areas with different possible village selections were generated using covariate-constrained randomization to balance baseline village-level population size, distance to the nearest health centre offering vaccinations, and MCV1 coverage among children aged 12-24 months. Randomization was restricted to assignments with standardized mean differences $\leq$0.20. The empirically estimated intracluster correlation coefficient (ICC) and the World Health Organization (WHO) recommended ICC values of 1/3 and 1/6 were considered for scenario analyses. The desired operating characteristics were 80% power at 0.05 one-sided type I error rate.

**Results:** Conventional calculations showed that target power for a realistic treatment effect could not be achieved with WHO recommended ICC values. Conventional calculations also showed a plateau in power after a certain cluster size. Our simulations could better match the actual design of the OptiMAx-Chad study with covariate adjustment and random village selection and showed that power did not plateau. Instead, power increased with increasing cluster size.

**Conclusion:** Planning complex cRCTs with covariate constrained randomization and a multi-nested data structure with conventional closed-form formulas can be misleading. Simulations, when matched to the intended design, can improve the planning of cRCTs.

**Funding:** This work was supported by the Gates Foundation, Seattle, WA USA.

## 2    Introduction

Cluster randomized clinical trials (cRCTs) are an important methodological tool for global health. In cRCTs, clusters, which may refer to structured groups of individuals or health facilities, are randomly assigned to study interventions to evaluate intervention effectiveness, often with the goal of estimating population-level effects.[1,2] Randomization can reduce selection bias and provides a sound statistical basis for establishing causality.[3,4] For population-based interventions, randomizing interventions at the cluster-level, rather than the individual-level, can reduce treatment contamination between study arms.[5,6] Provided many clusters are available, cluster randomization, on average, will produce study groups that are comparable on measured and unmeasured prognostic factors.[7] When few clusters are available, cRCTs are at risk of imbalanced prognostic factors. In these situations, covariate-constrained randomization is recommended.[8,9] Covariate-constrained randomization is a form of restricted randomization where the randomization schemes are limited to those that meet pre-specified balancing metrics.

Given the complexities involved in covariate-constrained randomization, using conventional closed-form formulas to design cRCTs may be suboptimal. Existing formulas for sample size and power for cRCTs assume complete balance and often involve a naïve (i.e., unadjusted) analysis.[10,11] These formulas cannot easily account for covariate-constrained randomization which requires statistical adjustment of balancing covariates in the analysis.[11] Furthermore, most formulas are limited to conventional cRCTs with participants nested within clusters (i.e., a two-level structure), and where the unit of randomization is the same as the unit of outcome collection. However, there are cRCTs with a multi-nested data structure, where the unit of randomization (e.g., health areas) differs from the unit of outcome collection (e.g., villages).[12,13]

Instead of closed-form calculations, adopting simulation-guided design for cRCTs may be beneficial.[2,14] Simulations are useful for optimizing trial design as they allow comparison of competing options, including statistical analyses and design choices. To optimize clinical development, this approach is increasingly being used to plan individually randomized trials, especially those that involve complex, innovative designs.[2,15,16] However, the use of simulations for cRCTs has been limited.[2]

Here, we demonstrate the utility of simulation-guided design for cRCTs using a real study called OptiMAx-Chad. We compared the estimated operating characteristics from a commonly used conventional closed-form formula and from simulations.[17]

## 3 Methods

### 3.1 Overview of OptiMAx-Chad Trial Design

OptiMAx-Chad is a two-arm cRCT evaluating the effectiveness of small-quantity lipid-based nutrient supplements (SQ-LNS) coupled with an existing Expanded Program on Immunization (EPI) for increasing childhood vaccine coverage in rural and remote villages in Ngouri, Chad. The study area comprises 313 villages across 12 health zones.

The primary endpoint of OptiMAx-Chad is the proportion of children aged 12-24 months with at least one dose of measles-containing vaccine (MCV1). Village-level cluster randomization was determined to be operationally infeasible. As a result, health zone, a higher geographic level in which villages are nested, was chosen as the unit of randomization. For outcome collection, as it was not possible to survey all villages, a random subset of villages will be selected. The result is a multi-level structure with randomization at the health zone level and outcome collection at the village level (Figure 1).

To prevent potential imbalance arising from the small number of clusters randomized, we applied covariate-constrained randomization with standardized mean differences (SMDs) of 0.20 at the village-level as the balancing metric.[18] A population-level baseline census survey conducted in the study area between December 2024 to January 2025 was used for randomization and random selection. The survey collected vaccination status of all children aged 12-24 months. The list of pre-specified baseline covariates for balance included village-level estimates of: population size; distance to the nearest health centre offering vaccinations; and MCV1 coverage among study eligible children. Village-level covariates were selected in consultation with the ALIMA implementation team using causal reasoning via directed acyclic graphs.

### 3.2 Analysis of Baseline Census Data

We calculated population-level descriptive statistics. We fit a logistic regression to model the baseline village-level MCV1 rates of children aged 12-24 months (the primary endpoint) using village population size and village distance to the nearest health centre providing vaccination as covariates. The coefficient estimates informed the coefficient values for our simulation study. From the baseline census data we obtained an estimate of the ICC at the unit of randomization (health-zone level; $ICC_H$) and outcome collection (village-level; $ICC_V$). Calculations are provided in supplementary material 1.

### 3.3 Closed-Form Formula for Power

We used a common closed-form power formula for a single-stage cRCT (supplementary material 1). We used the empirically estimated health-zone level ICC ($ICC_H$ = 0.048) and also considered $ICC_H$ values of 1/6 and 1/3 as recommended by the World Health Organization (WHO).[19] We considered varying, but approximately equal, cluster sizes. We calculated the average cluster size by multiplying the average number of eligible children per village (i.e., 14) in the census survey with the varying number of villages in the follow-up survey and then dividing the product by the number of clusters. We specified a one-sided alpha of 0.05 and assumed no loss to follow-up.

### 3.4 Simulation Study

Our simulations were planned using the ADEMP structure for pre-specification of simulation studies.[20] A detailed simulation protocol was prepared prior to conducting our simulation study. The simulation aimed to: determine the required number of villages to be sampled in the follow-up survey; characterize statistical power and type I error; and compare different analytical approaches for the OptiMAx-Chad primary analysis.

#### 3.4.1 Data Generating Mechanism

Our data generating mechanism was determined by analysis of the baseline census data. In our simulations, health zones and village selections were randomly assigned to each arm with covariate-constrained randomization using a Monte Carlo approximation to generate the list of possible assignments. We generated 1,000,000 iterations of health area assignments to the intervention and control arms in a 1:1 allocation ratio. For each health area assignment, a possible selection of $n$ villages in each arm was generated with a 1:1 allocation ratio. Within each selection, the number of villages selected was approximately proportional to the total number of villages in the health area.

We considered a varying number of villages per arm (60, 70, 80, and 90 per arm). We estimated the power with alternative hypothesis scenarios that consisted of absolute improvements in village-level

MCV1 coverage rates of 0.10, 0.15, and 0.20, and control event rates (CERs) varying from 0.55 to 0.75 at increments of 0.05. Type I error was characterized using the null hypothesis scenario, where we assumed there would be no improvements in coverage. The assumed beta coefficient values were empirically estimated from baseline data. The three values for the coefficients assumed the point estimate and the lower and upper 95% confidence intervals (supplementary material 1).

### 3.4.2 Estimands and Targets

The primary target was the null hypothesis of no difference between the outcomes in the control and treatment groups. Our estimand of interest was the treatment effect defined as the expected difference in the village-level MCV1 rate between the treatment and control groups among children aged 12-24 months at the 12-month follow-up survey. For the power calculation using the conventional formula, the targets were the same as above except the difference in treatment effect was defined at the health-zone level.

### 3.4.3 Candidate Analytic Methods

For our primary analysis, we considered quasi-binomial regression and beta-regression as candidate methods and compared them to a naïve Wald test (supplementary material 1).[21,22] Each analysis was calibrated to control the type I error rate at 0.05 (one-sided).

### 3.4.4 Performance Metrics

Our primary performance metrics were type I error and statistical power. This corresponded to the type I error rate when assuming no treatment effect (null scenario), and to statistical power when assuming alternative scenarios with varying treatment effects. We performed 10,000 simulations to characterize the type I error rate and 1,000 simulations to characterize the statistical power. We also considered Monte Carlo error (supplementary material 1).

## 3.5 Computation

We conducted our simulations using R version 4.4.0.[23] Packages used included *dplyr* and *tidyr* for data manipulation; *ggplot2* for data visualization; *lme4* for fitting the mixed effects logistic regression model; and *beta regression* for fitting the beta generalized linear model.[24-29]

### 3.6 Role of Funding Source

This publication is based on research funded by the Gates Foundation, which reviewed the study and provided non-binding recommendations to improve informativeness. The Foundation did not have a role in approving or writing the final protocol. The findings and conclusions contained within are those of the authors and do not necessarily reflect positions or policies of the Gates Foundation.

## 4 Results

### 4.1 Baseline Census Survey Results

Summary statistics for villages with at least five children aged 12-24 months in the baseline survey are presented in supplementary tables S1 and S2. Relevant covariates varied considerably by health area. For example, the mean village distance to the nearest health centre ranged from 1.2 km to 6.8 km, and the mean total village population ranged from 56 to 940 individuals. Baseline MCV1 coverage rates in villages also varied by health area, ranging from 0.66 to 0.99 (supplementary figure S1).

### 4.2 Power Calculations

Power calculations were performed with 12 health zones (6 per arm) being available for randomization. The calculated power to detect an absolute increase of 0.15 in MCV1 rate with varying cluster sizes is shown in Figure 2. Across all $ICC_H$ values and CERs, there was a plateau in power with the cluster size, where the power did not increase with any further increases in the cluster size reflecting the overall number of children surveyed for follow-up in each health zone. For CERs of 0.65 and 0.70, using the WHO recommended $ICC_H$ values of 1/3 and 1/6, the trial could not be powered adequately (e.g., 80% power) to detect an absolute increase in MCV1 rate of 0.15, which was considered an important difference for the OptiMAx program, irrespective of the cluster size. Under the empirically estimated $ICC_H$ values of 0.048, power improved; however, the trial could not meet the target power to detect a treatment effect of 0.15 under the CER of 0.65.

The relationship between power and treatment effect was further examined by fixing the cluster size.

Figure 3 shows the calculated statistical power at 90 villages per arm under the CER of 0.70 (our base case). Under the WHO recommended $ICC_H$ values, the trial could be powered above 80% but only to detect unrealistically large treatment effects such as perfect or near-perfect intervention performance. For example, under an $ICC_H$ of 1/3, complete vaccination coverage (i.e., vaccine coverage = 1.0) would be needed in the intervention arm to achieve statistical power greater than 80%. Alternative scenarios showed similar findings (supplementary figures S2-4).

### 4.3 Simulation-Based Trial Operating Characteristics

From our simulations, we noted that all analytic methods produced a slight inflation in the type I error rate at $\alpha$ = 0.05 as covariate-constrained randomization involves adjustment variables. To account for this, we opted to use $\alpha$ = 0.045 corresponding to a critical $Z$ value of 1.695 for all models.

In the base case, quasi-binomial regression showed inflated type I error rates, ranging from 0.083 to 0.089. (Table 1). Beta regression, however, controlled the type I error rate between 0.043 to 0.050. The naïve analysis showed similar control (type I error rate:0.044 to 0.053). A full breakdown of type I error by number of villages per arm, $ICC_V$, absolute increase in vaccination rate, and CER is available in supplementary material 1.

Both beta regression and the naïve analyses achieved 80% power to detect an absolute increase of 0.13 in the MCV1 rate with a sample size of 80 villages per trial arm. In general, the naïve analysis achieved slightly greater power than beta regression (Table 1). Lowering the CER decreased power, and increasing the $ICC_V$ to 1/3 also decreased the power, which is consistent with trends observed in the conventional power calculation. Importantly, the simulation results did not show an observable plateau in power when increasing the sub-cluster size (villages per arm; Figure 4). This is due to the fact the simulations better accounted for the village-level variabilities in the event rate and prognostic factor distribution. Alternative scenarios for the simulations are presented in supplementary figures S5-S12.

### 5 Discussion

In this study we compared simulation-guided design of a multi-nested cRCT using covariate-constrained randomization with conventional power calculations. In the absence of a reliable estimate of the ICC, it is a common practice to refer to the WHO's vaccination coverage survey guidance, which recommends using an ICC value between 1/6 and 1/3 for immunization surveys.[30] These $ICC_H$ values were considerably higher than what we empirically estimated using census survey data. Using these guesstimates of ICC, conventional power calculations showed the trial could only be powered above 80% if we assumed unrealistic treatment effects. While feasibility improved with the empirically estimated $ICC_H$ of 0.048, the results were still misleading: they showed a plateau in the relationship between power and number of children surveyed peri cluster (i.e., cluster size). In contrast, our simulations, more accurately captured the dynamic randomization procedure and village-level variabilities in covariates and prognostic factors, showed no plateau in the relationship between cluster size and statistical power. If design relied on power calculations, an inappropriate trial design could have been selected. For example, the power calculations plateaued well before the cluster size of 500(Figure 2. Assuming an average of 14 eligible children per village, this translates to ~36 villages per arm, resulting in an underpowered trial.

Simulations offer multiple advantages over conventional closed-form formulas. Our findings show that these formulas can be misleading for trial planning, even when influential assumptions (e.g., CER, ICC, and average cluster size) can be empirically estimated. This is because conventional calculations use point estimates, which do not fully capture variability, as inputs.[11,31] For example, cluster sizes cannot be adequately explained using a single point estimate, even if a standard deviation reflecting that variability is also used. In OptiMAx-Chad, we saw large variation in the number of eligible children at the sub-cluster level. Given the availability of population-level estimates from baseline data, our simulations were chosen to randomly sample based on the entire distribution of the collected data without imposing any explicit assumptions on the data generating mechanism. With conventional formulas, it is also difficult to account for nested data structures, where the units of randomization and outcome collection differ.[11] To calculate the power, we had to use the average sub-cluster size and the number of sub-clusters to estimate the overall cluster size, essentially ignoring the observed sub-cluster variability.

Simulations allow for comparison of different candidate statistical analyses. Adjustment of balancing covariates is recommended to avoid inflating the type I error rate.[8,9] The power calculation formula, which was based on a naïve analysis with a Wald test, assumes the trial design based on the power calculations can control the type I error rate at 0.05 (one-sided). However, simulations showed the naïve analysis with covariate-constrained randomization could not control the type I error rate with the significance level ($\alpha$) being set at 0.05. Controlling the type I error rate required a stricter $\alpha$. Our simulations allowed for comparison of two additional estimators: beta regression and quasi-binomial regression, both with covariate adjustment.

Our study demonstrates the benefits of adopting simulation-guided design in planning cRCTs with covariate-constrained randomization, an important tool for achieving balance when only a few clusters can be randomized.[8] Closed-form calculations cannot easily accommodate the design of cRCTs with covariate-constrained randomization, especially given a multi-level nested data structure.. Simulation-guided design offers an important analytical framework to integrate real-world complexity into cRCT planning. While we explored a multi-nested data structure in the OptiMAx-Chad, the benefits of simulation-guided design likely extend to planning conventional cRCTs, where the units of randomization and outcome collection match. Our results suggest that simulation-guided design can optimize cRCT planning by enabling more accurate estimation of operating characteristics compared to conventional trial planning.

There are strengths and limitations to our study. The strength of this work is bolstered by the involvement of a multi-disciplinary study team. We worked with clinical and implementation experts to ensure our simulations reflected valid assumptions, and we maximized the usefulness of baseline data in our simulations by randomly drawing the parameters directly from the data to accurately reflect their observed variation. We also explored the impact of several analytic models and compared their trade-offs before selecting the optimal estimator for our study.

This work should be considered in the contexts of its limitations. Our simulations were grounded in baseline census survey data which allowed us to be confident in our assessment of relevant confounders and balancing covariates; such comprehensive baseline data is uncommon and very resource intensive to generate. In its absence, however, other data, such as demographic health surveys coupled with subject-matter expertise, can be used to inform selection of balancing covariates. While our study shows that simulation-guided design is feasible for cRCTs even when considering covariate-constrained randomization, the simulations are resource intensive. In addition to expertise in statistical simulation and clinical trial design, there are substantial computational demands which can be costly in both finances and time. For example, we adopted Monte-Carlo probabilistic based random selection with 1,000,000 different iterations with varying treatment allocations. This was repeated for 10,000 simulations to characterize the type I error rate and 1,000 simulations under the alterative scenarios. Our simulation study required several days to complete even using specialized cloud computation services. While these barriers may limit the feasibility of simulation-guided design, advances in parallel computing and GPU-based infrastructure are making such simulations increasingly practical. Furthermore, the advantages offered in terms of improved trial efficiency can offset these costs.

## 6    Conclusion

Simulation-guided design of cRCTs can overcome the structural limitations of closed-form formulas for power and sample size calculations. To plan OptiMAx-Chad with only a small number of clusters for randomization, we leveraged comprehensive baseline data and conducted simulations to better estimate trial operating characteristics and compare multiple candidate estimators. Compared to closed-form formulas, our simulation study allowed different trade-offs in terms of sub-cluster size and error rates to be considered even with a multi-nested data structure and a dynamic randomization scheme. Our work demonstrates that simulation-guided design is a powerful planning tool for cRCTs.

## 7 Declaration of Interests

The authors have no conflicts of interest to declare.

## 8 Author Contributions

JJHP: contributed to the conceptualization and design of the study, conducted analyses, drafted the initial manuscript and provided revisions; RKM: contributed to the conceptualization and design of the study, drafted the initial manuscript and provided revisions. ND: conducted analyses, drafted the initial manuscript, and provided revisions; YY: contributed to data management and processing and provided revisions to the manuscript. SA contributed to manuscript drafting and revisions; KP and SS contributed to the conceptualization and design of this work and provided oversight and revisions to the manuscript.

## 9 Data Sharing Statement

Requests for data can be submitted to the corresponding author and will be considered by the author group in conjunction with ALIMA's leadership in Chad.

## 11   Tables

**Table 1: Comparison of operating characteristics between conventional closed-form formulas and multi-nested simulations based on empirically estimated ICC ($ICC_V = 0.24$ vs $ICC_H = 0.048$)**

| Villages per arm | Beta regression ($ICC_V = 0.24$) | Naïve ($ICC_V = 0.24$) | Conventional Formula ($ICC_H = 0.048$) |
|---|---|---|---|
| **Type I error rate at 0.70 CER** | | | |
| 60 | 0.0430 | 0.0457 | 0.05 (assumed) |
| 70 | 0.0483 | 0.0528 | 0.05 (assumed) |
| 80 | 0.0453 | 0.0485 | 0.05 (assumed) |
| 90 | 0.0500 | 0.0442 | 0.05 (assumed) |
| **Statistical power to detect 0.15 absolute improvement (0.70 CER)** | | | |
| 60 | 0.878 | 0.924 | 0.794 |
| 70 | 0.886 | 0.95 | 0.801 |
| 80 | 0.93 | 0.961 | 0.805 |
| 90 | 0.95 | 0.985 | 0.809 |

**Acronyms**: CER – control event rate; $ICC_V$ – village-level intracluster correlation coefficient; $ICC_H$ – health centre-level intracluster correlation coefficient

## 12 Figures

**Figure 1: Multi-level data structure of OptiMAx-Chad study**

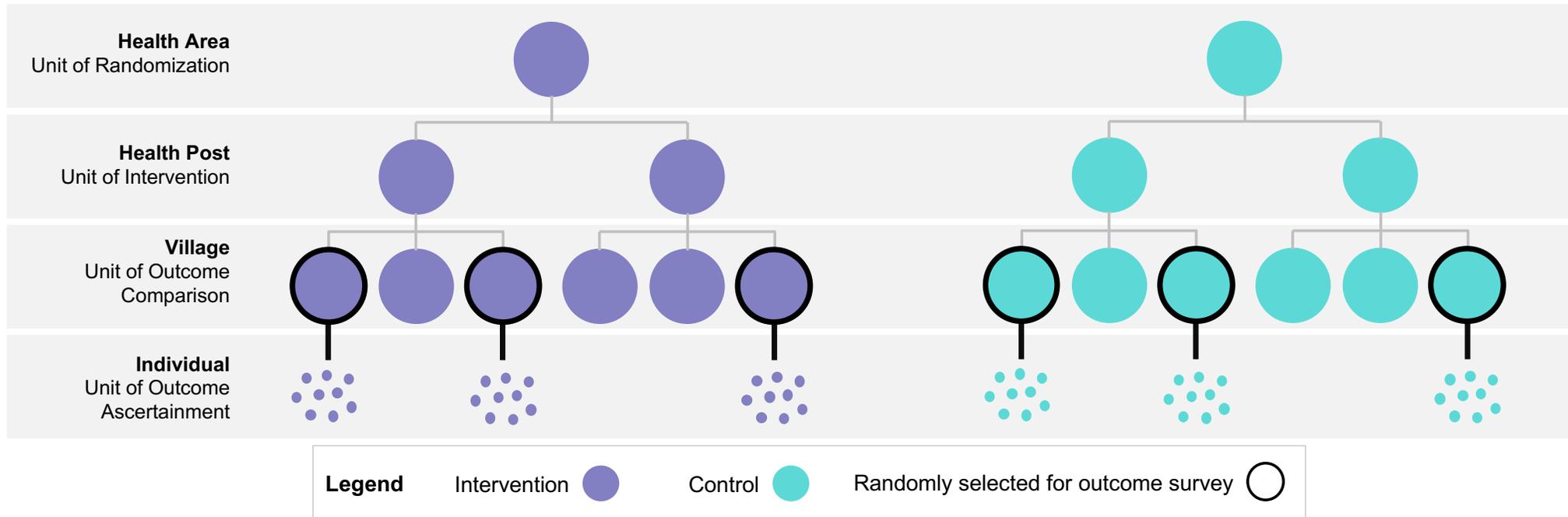

**Figure 2: Calculated power to detect an absolute increase of 0.15 in vaccine coverage rate with conventional formula with varying cluster size**

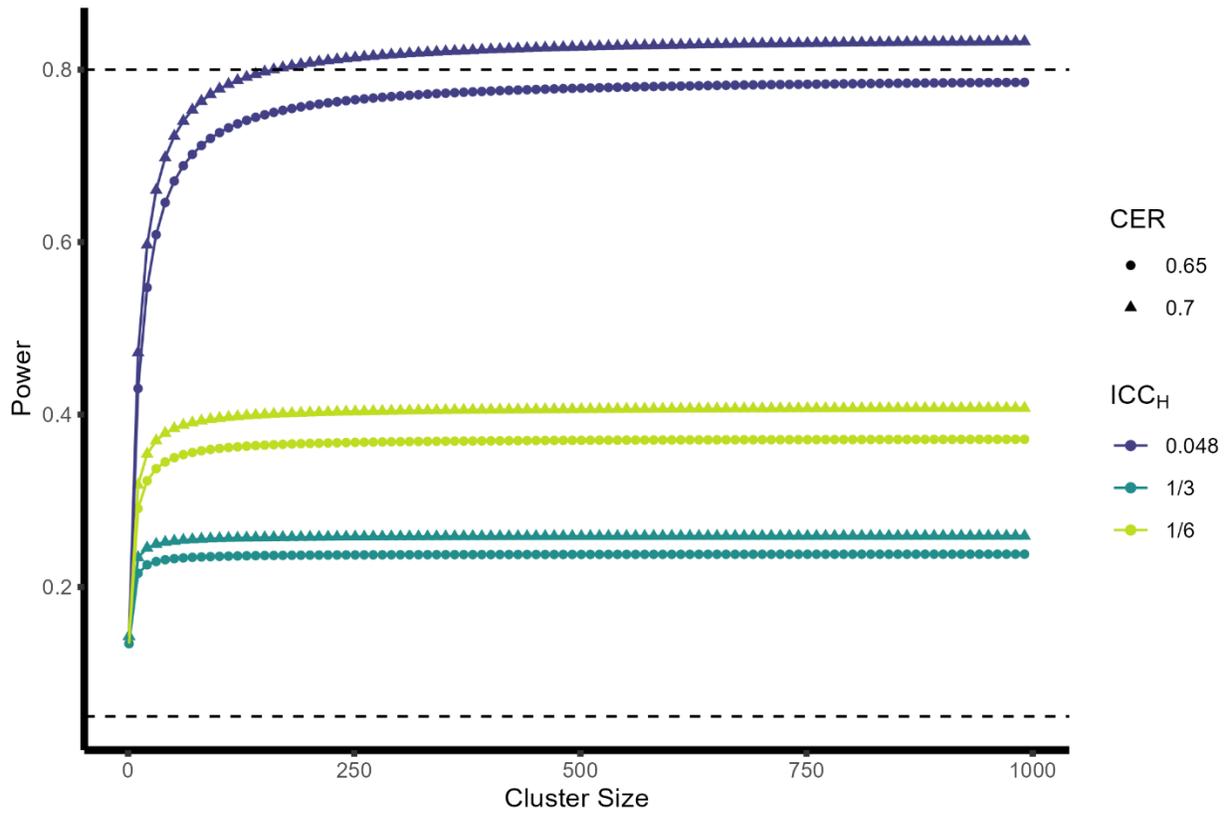

**Figure 3: Calculated statistical power with varying treatment effects for 90 villages per arm using conventional closed-form formula (CER = 0.70)**

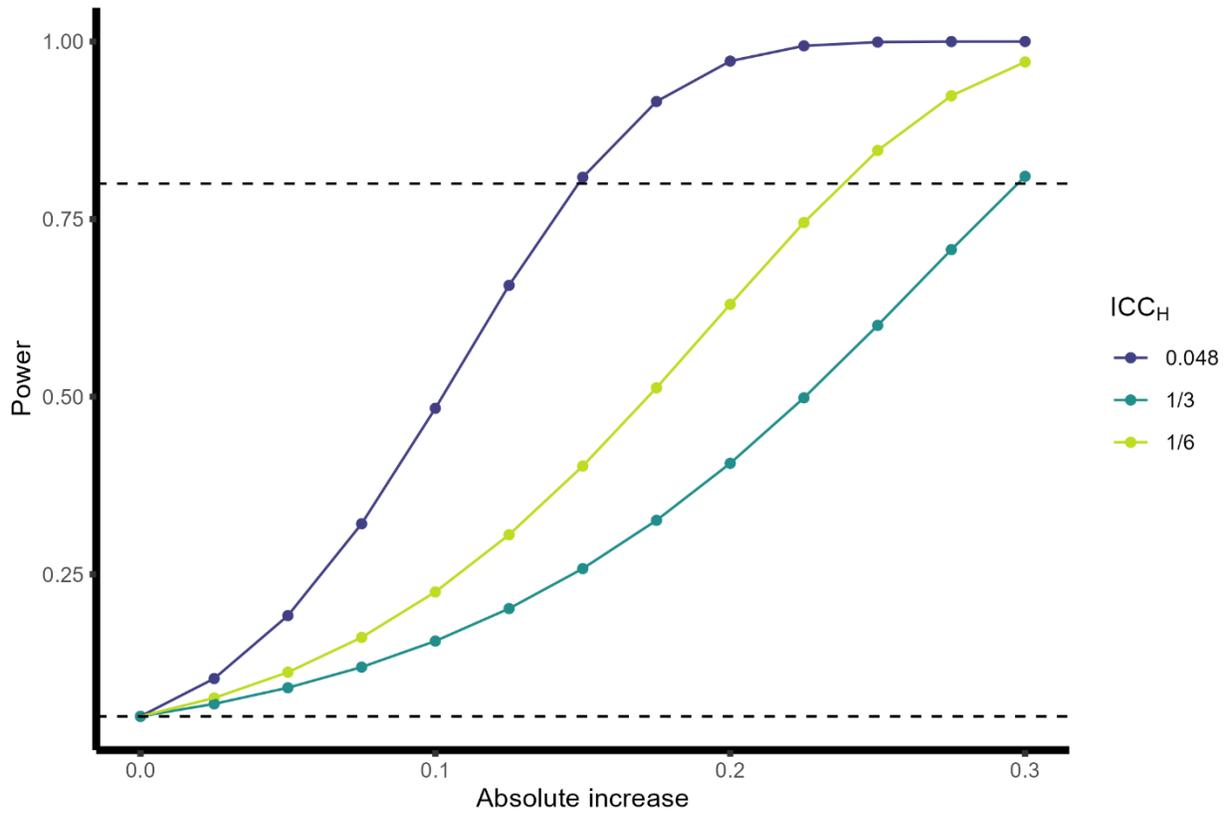

**Figure 4: Power of beta regression by absolute increase in vaccination rate with empirically estimated $ICC_v$ (CER=0.70)**

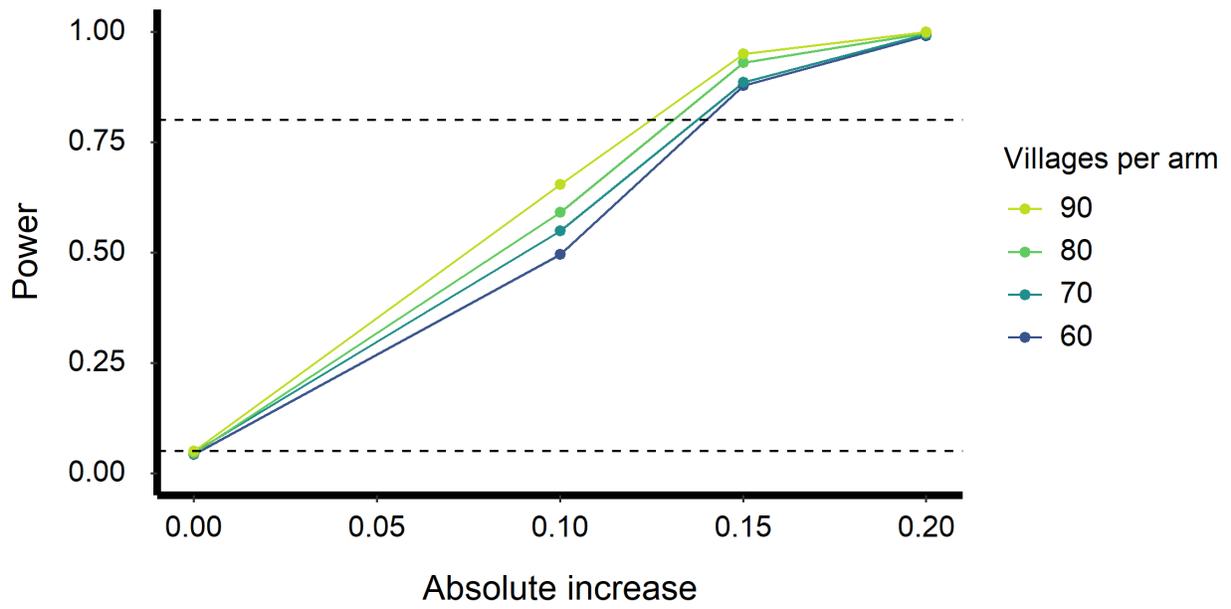

**Supplementary Materials 1:**

Supplementary to "*Comparison of Simulation-Guided Design to Closed-Form Power Calculations in Planning a Cluster Randomized Trial with Covariate-Constrained Randomization: A Case Study of Mass Small-Quantity Lipid Nutrient Supplementation Combined with an Expanded Program on Immunization in Rural Chad*"

# 1. Supplementary Methods

**Details of Baseline Census Data Analysis for Simulation Parameters including the Intracluster Correlation Coefficient (ICC) Calculations**

Let $Y_j^0$ be the total number of children aged 12-24 months with MCV1 and let $m_j^0$ be the total number of children aged 12-24 months in village $j$. Let $p_j$ be the population at the baseline survey, and $d_j$ be the distance to the nearest health center, then the logistic regression model is

$$Y_j^0 \sim Bin(m_j^0, \pi'_j),$$

$$\log \frac{\pi'_j}{1-\pi'_j} = \eta_0 + \eta_1 p_j + \eta_2 d_j.$$

The coefficient estimates from the fitted models were used to inform the coefficient values for our simulation study. We also used the baseline census-like data to obtain an estimate of the ICC at the village level (ICCV) using the following model

$$Y_j^0 \mid v_j \sim Bin(m_j^0, \pi'_j),$$

$$\log \frac{\pi'_j}{1-\pi'_j} = \eta_0 + v_j + \eta_1 p_j + \eta_2 d_j, v_j \sim N(0, \tau^2),$$

where $p_j$ and $d_j$ are the population and distance of village $j$ respectively. To avoid convergence issues, we standardize $p_j$ into $\tilde{p}_j$

$$\tilde{p}_j = \frac{p_j - \bar{p}}{sd(p)},$$

where $\bar{p}$ and $sd(p)$ were the mean and standard deviation of total village population. Lastly, to estimate the ICCv for binary outcomes,[1] we estimate it using the variance of the village level random effect

$$ICC_V = \frac{\hat{\tau}^2}{\frac{\pi^2}{3} + \hat{\tau}^2},$$

where $\pi^2/3$ is the residual variance of the logistic distribution.

The ICC for the health-zone cluster (denoted as ICC$_H$) was calculated by fitting the same mixed effects model as above but replacing $v_j \sim N(0, \tau^2)$ with $\delta_c \sim N(0, \omega^2)$, and the random effect for health zone with $c = 1, \ldots, 12$.

**Closed-form Formula for Power Calculations**

We revised the closed-form formula for sample size calculation for power calculation from Hayes and Moulton 2017.[2] This formula is for a single-stage cluster RCT that assumes equal cluster sizes based on the ICC for a one-sided difference in study arm proportions. The formula for the number of clusters required per study arm was rearranged as below:

$$Z_{1-\beta} = \sqrt{\frac{m(c-1)(\pi_0 - \pi_1)^2}{[\pi_0(1-\pi_0) + \pi_1(1-\pi_1)](1 + (m-1)ICC)}} - Z_{1-\alpha}$$

$$\text{Power} = \Phi(Z_{1-\beta})$$

**Simulation Details**

In each simulation replicate, a random pair of health area and village with an average village-level SMD not more than 0.2 was chosen. The number of vaccinated children aged 12-24 months that would be available for the follow-up survey were generated according to the health area assignment of this pair. Subsequently, the data were generated via a mixed-effects logistic regression. Let $Y_j^t$ be the number of children aged 12-24 months with MCV1 in village $j$ at time $t$, where $t = 0$ is the baseline census, and $t = 1$ is the follow-up survey. Then the $Y_j^1$ was generated as follows:

$$Y_j^1 \sim Bin(m_j^1, \pi_j^1),$$

$$\log \frac{\pi_j^1}{1-\pi_j^1} = \beta_0 + \alpha_j + \log \frac{Y_j^0}{m_j^0 - Y_j^0} + \beta a_j + \beta_1 p_j + \beta_2 d_j, \alpha_j \sim N(0, \tau^2).$$

Within each repetition we varied the absolute increase of MCV1 ($\delta_a = 0, 0.1, 0.15, 0.2$); the follow-up MCV1 rate in the control arm ($\pi_0 = 0.55, 0.6, 0.65, 0.7, 0.75$); the number of villages sampled per arm ($n$); the main effect of village population ($\beta_1 = 0.000268, 0.000370, 0.0004966$); the main effect of distance to nearest health centre ($\beta_2 = -0.60630, -0.0867, -0.0345$); and the ICC (0.24, 1/3). The three values for the coefficients are the lower 95% confidence interval, point estimate, and upper 95% confidence interval values, respectively, for each parameter. The first ICC$_V$ value was estimated from the baseline survey data and the second was the conservative recommendation for the World Health Organization's vaccination survey planning guidance.[3] The follow-up MCV1 rates in the control arm were based on expert opinion.

We also further defined coefficient set $i$ as $i^{th}$ value of $\beta_1$ and $\beta_2$. The combination of all these coefficient sets corresponded to 360 scenarios per repetition. We defined the base-case simulation scenario as: a MCV1 rate of 0.70 in the control arm; village population coefficient of 0.0002668; distance to nearest health centre coefficient of -0.0606; and an ICC$_V$ of 0.24 while varying the number of villages sampled per arm.

*Candidate Analyses*

In our simulations, we considered quasi-binomial regression and beta regression for our primary analysis.[4,5] We included interaction terms between $p_j$ and $d_j$ to express our uncertainty about the true DGM. We compared these two methods and a naïve Wald test without any covariate adjustments. The quasi-binomial was modelled as follows:

$$\mathbb{E} Y_j^1 = m_j^1 \pi_j (1 - \pi_j)$$

$$Var(Y_j^1) = \varphi m_j^1 \pi_j (1 - \pi_j)$$

$$\log \frac{\pi_j}{1-\pi_j} = \beta_0 + \beta a_j + \beta_1 \frac{Y_j^0}{m_j^0} + \beta_2 p_j + \beta_3 d_j + \beta_4 p_j d_j.$$

The beta regression model was fit with the mean-precision parameterization

$$\frac{Y_j^1}{m_j^1} \sim Beta(\pi_j, \varphi)$$

$$Var(Y_j^1) = \varphi m_j^1 \pi_j (1 - \pi_j)$$

$$\log \frac{\pi_j}{1-\pi_j} = \beta_0 + \beta a_j + \beta_1 \frac{Y_j^0}{m_j^0} + \beta_2 p_j + \beta_3 d_j + \beta_4 p_j d_j.$$

To avoid having proportions exact at 0 or 1, $\pi_j$ was transformed to lie strictly within the (0,1) boundary.

and for each model we were testing the hypothesis

$$H_0: \beta = 0 \text{ vs. } H_1: \beta > 0$$

at the significance level of 0.05 by using the Wald test with the nominal point estimate and standard errors.

**Monte Carlo Error**

Any simulation study involves uncertainty in our estimates in using a finite number of replications. When estimating the Type 1 error Power, we utilized 10,000 and 1,000 repetitions respectively.

To quantify the error in our Monte Carlo simulation we use results for the Monte Carlo standard error.[6] Let $n_{rep}$ be the number of repetitions in our simulation and we assume the null hypothesis is true (assuming an absolute increase of vaccine coverage =0), then our point estimate for the type 1 error is:

$$\widehat{\text{Type 1 error}} = \frac{1}{n_{rep}} \sum_{k=1}^{n_{rep}} I(p_k \leq \alpha)$$

with the Monte Carlo Standard Error of

$$\widehat{MCSE}_{\widehat{\text{Type 1 error}}} = \sqrt{\frac{\widehat{\text{Type 1 error}}(1 - \widehat{\text{Type 1 error}})}{n_{rep}}}$$

Thus, using the central limit theorem, we then arrive at

$$\widehat{\text{Type 1 error}} \sim \text{Normal}\left(\text{Type 1 error}, \frac{\widehat{\text{Type 1 error}}(1 - \widehat{\text{Type 1 error}})}{n_{rep}}\right)$$

and we can then construct the corresponding 95% Monte Carlo confidence intervals. We may also construct the same confidence interval for power by setting the absolute increase in vaccination rate $> 0$ and using the same formulas.

## 2. Supplementary for Baseline Census Data

*Supplementary Table S1. Descriptive statistics of villages in Ngouri, Chad by health area for Amerom, Blachidi, Hagrerom, Kalimba, and Kindjira*

| Characteristic | Amerom N of villages = 37 | Blachidi N of villages = 28 | Boulorom N of villages = 16 | Hagrerom N of villages = 19 | Kalimba N of villages = 17 | Kindjira N of villages = 15 |
|---|---|---|---|---|---|---|
| Mean (SD) of village distance to nearest health center[1] | 6.4 (3.3) | 6.8 (4.2) | 6.3 (4.5) | 4.6 (2.8) | 4.3 (1.7) | 2.7 (1.2) |
| Mean (SD) of total village population size[2] | 198.5 (267.0) | 56.0 (69.8) | 292.7 (260.7) | 276.7 (287.4) | 233.5 (189.0) | 302.8 (414.7) |
| Mean (SD) of number of children aged 12-24 months in village[3] | 18.6 (26.9) | 11.3 (7.5) | 9.7 (5.4) | 14.2 (9.1) | 12.2 (6.8) | 11.1 (8.3) |
| Total number of children aged 12-24 months in group[4] | 688 | 316 | 155 | 269 | 207 | 167 |
| Total number (proportion) of children aged 12-24 months with MCV1 in group[5] | 534 (0.78) | 199 (0.63) | 110 (0.71) | 208 (0.77) | 143 (0.69) | 151 (0.90) |
| Mean (SD) of village MCV1 rate[6] | 0.73 (0.20) | 0.61 (0.23) | 0.70 (0.36) | 0.75 (0.23) | 0.70 (0.19) | 0.87 (0.20) |
| Median (Q1, Q3) of village MCV1 rate | 0.73 (0.60, 0.92) | 0.65 (0.39, 0.78) | 0.89 (0.38, 1.00) | 0.80 (0.71, 0.85) | 0.70 (0.60, 0.86) | 1.00 (0.78, 1.00) |

1 Distance of a village to the nearest health centre, derived from GPS information during data collection

2 Total population of a village, derived from census data

3 Number of children aged 12-24 months in a village as recorded in baseline survey

4 Sum of number of children aged 12-24 months across all villages in each health area

5 Sum of number of children aged 12-24 months with MCV1 across all villages in each health area. The proportion is the number of children aged 12-24 months with MCV1 vaccination divided by the number of children aged 12-24 months in total in each health area

6 Number of children aged 12-24 months with MCV1 in a village divided by number of children aged 12-24 months in the village, as recorded in baseline survey

*Supplementary Table S2. Descriptive statistics of villages in Ngouri, Chad by health area for Kournotoulo, Loulou Kmaerom, Madem, Matoura, Safaye and Zingui*

| Characteristic | Kournotoulo N of villages = 11 | Loulou Kamerom N of villages = 10 | Madem N of villages = 10 | Matoura N of villages = 14 | Safaye N of villages = 17 | Zingui N of villages = 6 |
|---|---|---|---|---|---|---|
| Mean (SD) of village distance to nearest health center[1] | 2.93 (1.41) | 3.70 (2.40) | 2.82 (1.94) | 4.16 (2.19) | 4.06 (2.49) | 1.17 (0.60) |
| Mean (SD) of total village population size[2] | 467.7 (217.6) | 302.9 (125.4) | 312.5 (291.4) | 333.9 (236.1) | 286.5 (278.8) | 940.3 (404.7) |
| Mean (SD) of number of children aged 12-24 months in village[3] | 18.5 (11.2) | 11.8 (4.3) | 14.8 (9.5) | 15.8 (11.1) | 13.4 (7.2) | 29.0 (17.1) |
| Total number of children aged 12-24 months in group[4] | 203 | 118 | 148 | 221 | 228 | 174 |
| Total number (proportion) of children aged 12-24 months with MCV1 in group[5] | 154 (0.76) | 84 (0.71) | 97 (0.66) | 188 (0.85) | 159 (0.70) | 115 (0.66) |
| Mean (SD) of village MCV1 rate[6] | 0.75 (0.16) | 0.74 (0.20) | 0.66 (0.20) | 0.83 (0.19) | 0.68 (0.20) | 0.64 (0.16) |
| Median (Q1, Q3) of village MCV1 rate | 0.76 (0.73, 0.84) | 0.75 (0.67, 0.83) | 0.62 (0.60, 0.78) | 0.89 (0.81, 0.99) | 0.70 (0.60, 0.77) | 0.65 (0.59, 0.75) |

[1]Distance of a village to the nearest health centre, derived from GPS information during data collection

[2]Total population of a village, derived from census data

[3]Number of children aged 12-24 months in a village as recorded in baseline survey

[4]Sum of number of children aged 12-24 months across all villages in each health area

[5]Sum of number of children aged 12-24 months with MCV1 across all villages in each health area. The proportion is the number of children aged 12-24 months with MCV1 vaccination divided by the number of children aged 12-24 months in total in each health area

[6]Number of children aged 12-24 months with MCV1 in a village divided by number of children aged 12-24 months in the village, as recorded in baseline survey

*Supplementary Figure S1.   Density plot of measles 1 vaccine coverage across villages by health area*

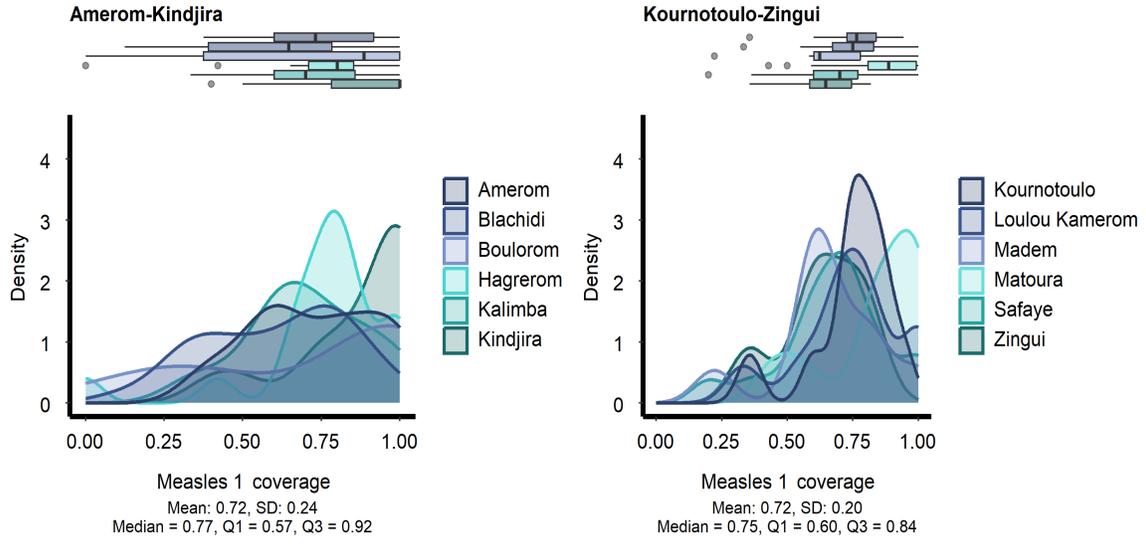

# 3. Supplementary for Closed-Form Power Calculations

*Supplementary Figure S2.*     *Power by absolute increase in vaccination rate (CER = 0.65)*

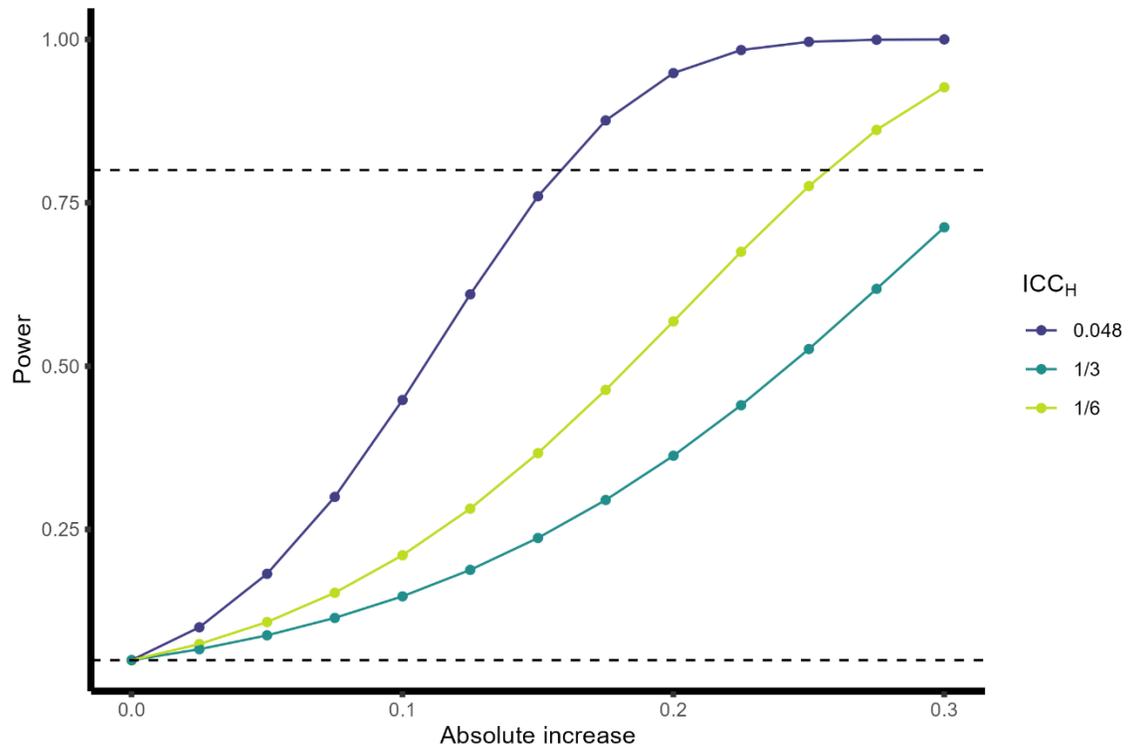

*Supplementary Figure S3.     Power by absolute increase in vaccination rate (CER = 0.6)*

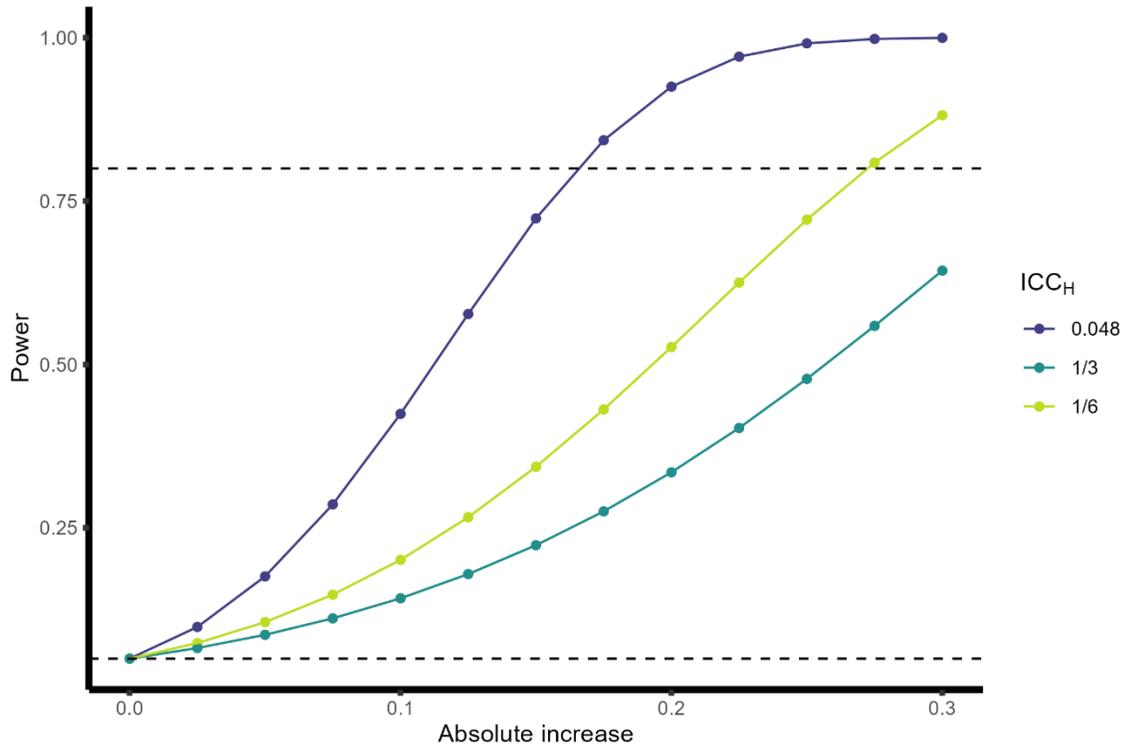

*Supplementary Figure S4.*     *Power by absolute increase in vaccination rate (CER = 0.55)*

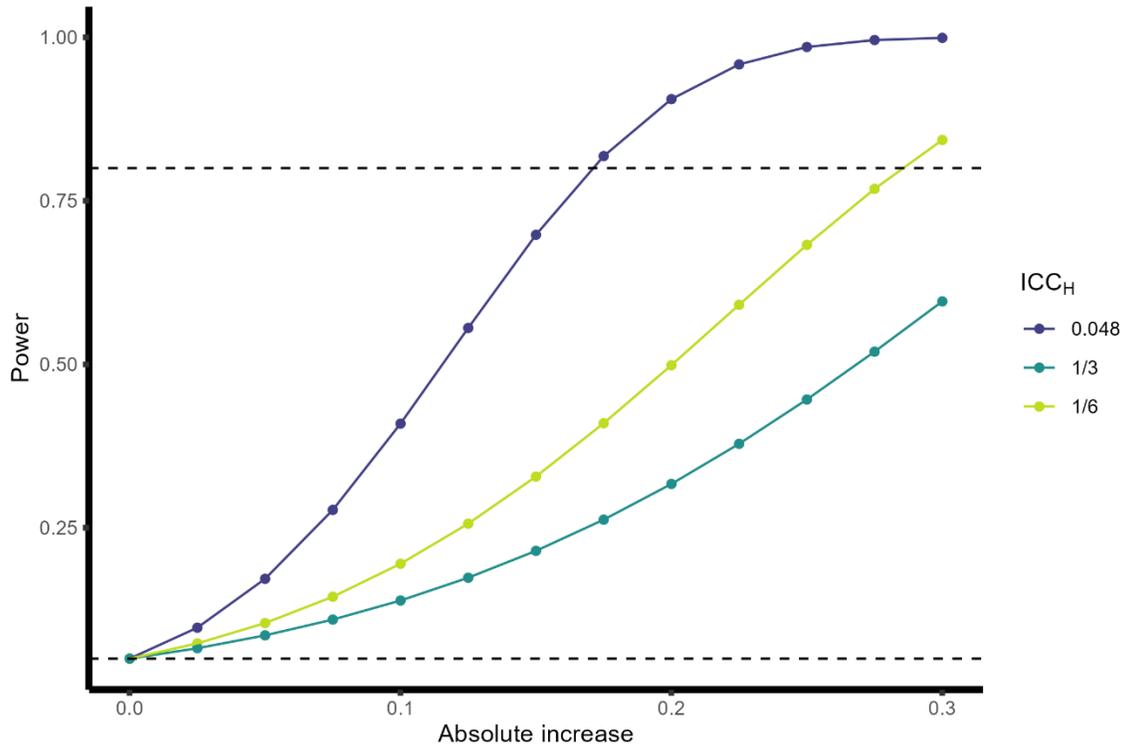

## 4. Supplementary for Type I Error Rates from Simulations

*Supplementary Table S3. Estimates and 95% confidence intervals of type I error rate of different methods under the base case (MCV1 rate of 0.70 in the control arm, coefficient set 1, ICC$_v$ of 0.24)*

| Villages per arm | Quasi-binomial | Beta | Naive |
|---|---|---|---|
| 60 | 0.083 (0.077, 0.088) | 0.043 (0.039, 0.047) | 0.046 (0.042, 0.050) |
| 70 | 0.088 (0.083, 0.094) | 0.048 (0.044, 0.052) | 0.053 (0.048, 0.057) |
| 80 | 0.089 (0.084, 0.095) | 0.045 (0.041, 0.049) | 0.048 (0.044, 0.053) |
| 90 | 0.089 (0.084, 0.095) | 0.050 (0.046, 0.054) | 0.044 (0.040, 0.048) |

*Supplementary Table S4. Estimates and 95% confidence intervals of type I error rate of different methods: Control MCV1 rate: 0.55; Coefficient set: 1; ICC$_v$: 0.24*

| Villages per arm | Quasi-binomial | Beta | Naive |
|---|---|---|---|
| 60 | 0.086 (0.081, 0.092) | 0.049 (0.045, 0.053) | 0.056 (0.051, 0.060) |
| 70 | 0.088 (0.082, 0.093) | 0.049 (0.044, 0.053) | 0.057 (0.052, 0.061) |
| 80 | 0.096 (0.090, 0.101) | 0.054 (0.050, 0.059) | 0.062 (0.057, 0.067) |
| 90 | 0.096 (0.090, 0.102) | 0.064 (0.059, 0.068) | 0.59 .054, 0.064) |

*Supplementary Table S5. Estimates and 95% confidence intervals of type I error rate of different methods: Control MCV1 rate: 0.55; Coefficient set: 1; ICC$_v$: 1/3*

| Villages per arm | Quasi-binomial | Beta | Naive |
|---|---|---|---|
| 60 | 0.092 (0.086, 0.097) | 0.046 (0.042, 0.050) | 0.055 (0.050, 0.059) |
| 70 | 0.094 (0.088, 0.099) | 0.045 (0.041, 0.049) | 0.054 (0.049, 0.058) |
| 80 | 0.093 (0.087, 0.098) | 0.047 (0.043, 0.051) | 0.055 (0.050, 0.059) |
| 90 | 0.094 (0.088, 0.100) | 0.059 (0.055, 0.064) | 0.055 (0.051, 0.060) |

*Supplementary Table S6. Estimates and 95% confidence intervals of type I error rate of different methods: Control MCV1 rate: 0.55; Coefficient set: 2; $ICC_v$: 0.27*

| Villages per arm | Quasi-binomial | Beta | Naive |
|---|---|---|---|
| 60 | 0.086 (0.080, 0.091) | 0.051 (0.047, 0.055) | 0.055 (0.050, 0.059) |
| 70 | 0.086 (0.080, 0.091) | 0.049 (0.045, 0.053) | 0.057 (0.053, 0.062) |
| 80 | 0.092 (0.086, 0.097) | 0.050 (0.046, 0.054) | 0.056 (0.051, 0.060) |
| 90 | 0.091 (0.085, 0.097) | 0.064 (0.059, 0.068) | 0.056 (0.051, 0.060) |

*Supplementary Table S7. Estimates and 95% confidence intervals of type I error rate of different methods: Control MCV1 rate: 0.55; Coefficient set: 2; $ICC_v$: 1/3*

| Villages per arm | Quasi-binomial | Beta | Naive |
|---|---|---|---|
| 60 | 0.089 (0.084, 0.095) | 0.047 (0.043, 0.051) | 0.056 (0.051, 0.060) |
| 70 | 0.088 (0.083, 0.094) | 0.046 (0.042, 0.050) | 0.058 (0.053, 0.063) |
| 80 | 0.095 (0.089, 0.101) | 0.041 (0.037, 0.045) | 0.049 (0.045, 0.053) |
| 90 | 0.095 (0.089, 0.101) | 0.056 (0.051, 0.060) | 0.055 (0.050, 0.059) |

*Supplementary Table S8. Estimates and 95% confidence intervals of type I error rate of different methods: Control MCV1 rate: 0.55; Coefficient set: 3; $ICC_v$: 0.24*

| Villages per arm | Quasi-binomial | Beta | Naive |
|---|---|---|---|
| 60 | 0.087 (0.081, 0.092) | 0.050 (0.046, 0.055) | 0.053 (0.049, 0.058) |
| 70 | 0.090 (0.084, 0.096) | 0.054 (0.050, 0.058) | 0.061 (0.056, 0.066) |
| 80 | 0.095 (0.089, 0.100) | 0.055 (0.050, 0.059) | 0.060 (0.056, 0.065) |
| 90 | 0.089 (0.084, 0.095) | 0.070 (0.065, 0.075) | 0.063 (0.058, 0.068) |

Supplementary Table S9. *Estimates and 95% confidence intervals of type I error rate of different methods: Control MCV1 rate: 0.55; Coefficient set: 3; $ICC_V$: 1/3*

| Villages per arm | Quasi-binomial | Beta | Naive |
|---|---|---|---|
| 60 | 0.089 (0.083, 0.094) | 0.048 (0.044, 0.053) | 0.055 (0.050, 0.059) |
| 70 | 0.093 (0.087, 0.099) | 0.045 (0.041, 0.049) | 0.054 (0.050, 0.059) |
| 80 | 0.091 (0.086, 0.097) | 0.047 (0.042, 0.051) | 0.055 (0.050, 0.059) |
| 90 | 0.101 (0.095, 0.107) | 0.059 (0.055, 0.064) | 0.062 (0.057, 0.067) |

Supplementary Table S10. *Estimates and 95% confidence intervals of type I error rate of different methods: Control MCV1 rate: 0.60; Coefficient set: 1; $ICC_V$: 0.24*

| Villages per arm | Quasi-binomial | Beta | Naive |
|---|---|---|---|
| 60 | 0.086 (0.081, 0.092) | 0.051 (0.047, 0.056) | 0.058 (0.054, 0.063) |
| 70 | 0.088 (0.082, 0.093) | 0.048 (0.043, 0.052) | 0.059 (0.054, 0.064) |
| 80 | 0.087 (0.082, 0.093) | 0.047 (0.043, 0.051) | 0.053 (0.049, 0.057) |
| 90 | 0.092 (0.086, 0.098) | 0.063 (0.058, 0.068) | 0.053 (0.048, 0.057) |

*Supplementary Table S11.      Estimates and 95% confidence intervals of type I error rate of different methods: Control MCV1 rate: 0.60; Coefficient set: 1; $ICC_V$: 1/3*

| Villages per arm | Quasi-binomial | Beta | Naive |
|---|---|---|---|
| 60 | 0.092 (0.086, 0.097) | 0.043 (0.039, 0.047) | 0.053 (0.049, 0.057) |
| 70 | 0.092 (0.086, 0.098) | 0.045 (0.041, 0.049) | 0.053 (0.048, 0.057) |
| 80 | 0.099 (0.093, 0.105) | 0.045 (0.041, 0.049) | 0.054 (0.049, 0.058) |
| 90 | 0.097 (0.091, 0.103) | 0.053 (0.049, 0.057) | 0.054 (0.050, 0.059) |

*Supplementary Table S12.      Estimates and 95% confidence intervals of type I error rate of different methods: Control MCV1 rate: 0.60; Coefficient set: 2; $ICC_V$: 0.24*

| Villages per arm | Quasi-binomial | Beta | Naive |
|---|---|---|---|
| 60 | 0.084 (0.079, 0.090) | 0.050 (0.046, 0.054) | 0.056 (0.051, 0.060) |
| 70 | 0.084 (0.079, 0.090) | 0.049 (0.045, 0.054) | 0.058 (0.053, 0.062) |
| 80 | 0.090 (0.085, 0.096) | 0.050 (0.046, 0.054) | 0.057 (0.052, 0.061) |
| 90 | 0.094 (0.089, 0.100) | 0.055 (0.051, 0.060) | 0.048 (0.043, 0.052) |

*Supplementary Table S13.      Estimates and 95% confidence intervals of type I error rate of different methods: Control MCV1 rate: 0.60; Coefficient set: 2; $ICC_V$: 1/3*

| Villages per arm | Quasi-binomial | Beta | Naive |
|---|---|---|---|
| 60 | 0.096 (0.090, 0.102) | 0.050 (0.045, 0.054) | 0.054 (0.050, 0.059) |
| 70 | 0.086 (0.080, 0.091) | 0.040 (0.036, 0.044) | 0.051 (0.047, 0.055) |
| 80 | 0.096 (0.090, 0.102) | 0.042 (0.038, 0.046) | 0.048 (0.043, 0.052) |
| 90 | 0.092 (0.087, 0.098) | 0.049 (0.044, 0.053) | 0.048 (0.043, 0.052) |

*Supplementary Table S14.     Estimates and 95% confidence intervals of type I error rate of different methods: Control MCV1 rate: 0.60; Coefficient set: 3; $ICC_v$: 0.24*

| Villages per arm | Quasi-binomial | Beta | Naive |
|---|---|---|---|
| 60 | 0.089 (0.084, 0.095) | 0.054 (0.049, 0.058) | 0.054 (0.050, 0.059) |
| 70 | 0.093 (0.088, 0.099) | 0.049 (0.045, 0.053) | 0.055 (0.051, 0.060) |
| 80 | 0.088 (0.083, 0.094) | 0.049 (0.044, 0.053) | 0.056 (0.052, 0.061) |
| 90 | 0.087 (0.081, 0.092) | 0.058 (0.053, 0.063) | 0.052 (0.048, 0.057) |

*Supplementary Table S15.     Estimates and 95% confidence intervals of type I error rate of different methods: Control MCV1 rate: 0.60; Coefficient set: 3; $ICC_v$: 1/3*

| Villages per arm | Quasi-binomial | Beta | Naive |
|---|---|---|---|
| 60 | 0.087 (0.081, 0.092) | 0.044 (0.040, 0.049) | 0.050 (0.046, 0.055) |
| 70 | 0.088 (0.082, 0.094) | 0.042 (0.038, 0.046) | 0.051 (0.046, 0.055) |
| 80 | 0.096 (0.090, 0.102) | 0.047 (0.043, 0.051) | 0.052 (0.048, 0.056) |
| 90 | 0.092 (0.087, 0.098) | 0.054 (0.049, 0.058) | 0.053 (0.048, 0.057) |

*Supplementary Table S16.     Estimates and 95% confidence intervals of type I error rate of different methods: Control MCV1 rate: 0.65; Coefficient set: 1; $ICC_v$: 0.24*

| Villages per arm | Quasi-binomial | Beta | Naive |
|---|---|---|---|
| 60 | 0.082 (0.077, 0.087) | 0.048 (0.044, 0.052) | 0.052 (0.047, 0.056) |
| 70 | 0.090 (0.085, 0.096) | 0.049 (0.045, 0.053) | 0.055 (0.051, 0.059) |
| 80 | 0.087 (0.082, 0.093) | 0.049 (0.044, 0.053) | 0.052 (0.047, 0.056) |
| 90 | 0.087 (0.082, 0.093) | 0.052 (0.048, 0.056) | 0.048 (0.044, 0.053) |

*Supplementary Table S17.  Estimates and 95% confidence intervals of type I error rate of different methods: Control MCV1 rate: 0.65; Coefficient set: 1; $ICC_V$: 1/3*

| Villages per arm | Quasi-binomial | Beta | Naive |
|---|---|---|---|
| 60 | 0.089 (0.084, 0.095) | 0.042 (0.039, 0.046) | 0.053 (0.049, 0.057) |
| 70 | 0.090 (0.084, 0.095) | 0.043 (0.040, 0.047) | 0.052 (0.048, 0.057) |
| 80 | 0.093 (0.087, 0.099) | 0.043 (0.039, 0.047) | 0.050 (0.046, 0.054) |
| 90 | 0.094 (0.088, 0.100) | 0.050 (0.045, 0.054) | 0.049 (0.044, 0.053) |

*Supplementary Table S18.  Estimates and 95% confidence intervals of type I error rate of different methods: Control MCV1 rate: 0.65; Coefficient set: 2; $ICC_V$: 0.24*

| Villages per arm | Quasi-binomial | Beta | Naive |
|---|---|---|---|
| 60 | 0.086 (0.080, 0.091) | 0.049 (0.045, 0.053) | 0.054 (0.050, 0.059) |
| 70 | 0.090 (0.084, 0.096) | 0.042 (0.038, 0.046) | 0.054 (0.049, 0.058) |
| 80 | 0.086 (0.080, 0.091) | 0.046 (0.042, 0.051) | 0.052 (0.048, 0.056) |
| 90 | 0.086 (0.081, 0.092) | 0.053 (0.048, 0.057) | 0.43 .039, 0.047) |

*Supplementary Table S19.        Estimates and 95% confidence intervals of type I error rate of different methods: Control MCV1 rate: 0.65; Coefficient set: 2; $ICC_V$: 1/3*

| Villages per arm | Quasi-binomial | Beta | Naive |
| --- | --- | --- | --- |
| 60 | 0.093 (0.088, 0.099) | 0.049 (0.045, 0.053) | 0.054 (0.050, 0.059) |
| 70 | 0.092 (0.086, 0.097) | 0.040 (0.036, 0.044) | 0.050 (0.046, 0.054) |
| 80 | 0.092 (0.087, 0.098) | 0.038 (0.034, 0.042) | 0.047 (0.043, 0.052) |
| 90 | 0.087 (0.082, 0.093) | 0.046 (0.042, 0.050) | 0.044 (0.040, 0.049) |

*Supplementary Table S20.        Estimates and 95% confidence intervals of type I error rate of different methods: Control MCV1 rate: 0.65; Coefficient set: 3; $ICC_V$: 0.24*

| Villages per arm | Quasi-binomial | Beta | Naive |
| --- | --- | --- | --- |
| 60 | 0.088 (0.082, 0.093) | 0.050 (0.045, 0.054) | 0.049 (0.044, 0.053) |
| 70 | 0.088 (0.082, 0.093) | 0.047 (0.043, 0.051) | 0.053 (0.049, 0.058) |
| 80 | 0.091 (0.086, 0.097) | 0.047 (0.043, 0.051) | 0.053 (0.048, 0.057) |
| 90 | 0.089 (0.083, 0.094) | 0.052 (0.048, 0.056) | 0.046 (0.042, 0.050) |

*Supplementary Table S21.        Estimates and 95% confidence intervals of type I error rate of different methods: Control MCV1 rate: 0.65; Coefficient set: 3; $ICC_V$: 1/3*

| Villages per arm | Quasi-binomial | Beta | Naive |
| --- | --- | --- | --- |
| 60 | 0.086 (0.081, 0.092) | 0.041 (0.037, 0.045) | 0.050 (0.046, 0.054) |
| 70 | 0.094 (0.089, 0.100) | 0.043 (0.039, 0.047) | 0.053 (0.049, 0.058) |
| 80 | 0.091 (0.085, 0.097) | 0.040 (0.036, 0.044) | 0.049 (0.045, 0.053) |
| 90 | 0.089 (0.083, 0.094) | 0.045 (0.041, 0.049) | 0.050 (0.046, 0.054) |

*Supplementary Table S22.   Estimates and 95% confidence intervals of type I error rate of different methods: Control MCV1 rate: 0.7; Coefficient set: 1; $ICC_V$: 1/3*

| Villages per arm | Quasi-binomial | Beta | Naive |
|---|---|---|---|
| 60 | 0.092 (0.086, 0.097) | 0.046 (0.042, 0.050) | 0.052 (0.047, 0.056) |
| 70 | 0.087 (0.081, 0.092) | 0.039 (0.035, 0.043) | 0.046 (0.042, 0.050) |
| 80 | 0.093 (0.087, 0.098) | 0.042 (0.038, 0.046) | 0.052 (0.048, 0.056) |
| 90 | 0.093 (0.088, 0.099) | 0.045 (0.041, 0.049) | 0.048 (0.044, 0.052) |

*Supplementary Table S23.   Estimates and 95% confidence intervals of type I error rate of different methods: Control MCV1 rate: 0.7; Coefficient set: 2; $ICC_V$: 0.24*

| Villages per arm | Quasi-binomial | Beta | Naive |
|---|---|---|---|
| 60 | 0.089 (0.083, 0.094) | 0.049 (0.045, 0.053) | 0.055 (0.051, 0.060) |
| 70 | 0.092 (0.086, 0.097) | 0.049 (0.045, 0.053) | 0.054 (0.050, 0.059) |
| 80 | 0.092 (0.086, 0.097) | 0.046 (0.042, 0.050) | 0.053 (0.048, 0.057) |
| 90 | 0.092 (0.087, 0.098) | 0.045 (0.041, 0.049) | 0.042 (0.038, 0.046) |

*Supplementary Table S24.   Estimates and 95% confidence intervals of type I error rate of different methods: Control MCV1 rate: 0.7; Coefficient set: 2; $ICC_V$: 1/3*

| Villages per arm | Quasi-binomial | Beta | Naive |
| --- | --- | --- | --- |
| 60 | 0.084 (0.079, 0.090) | 0.043 (0.039, 0.047) | 0.047 (0.043, 0.051) |
| 70 | 0.087 (0.081, 0.092) | 0.047 (0.042, 0.051) | 0.051 (0.047, 0.056) |
| 80 | 0.090 (0.084, 0.095) | 0.046 (0.042, 0.050) | 0.052 (0.048, 0.057) |
| 90 | 0.089 (0.083, 0.095) | 0.051 (0.047, 0.055) | 0.044 (0.040, 0.048) |

*Supplementary Table S25.* *Estimates and 95% confidence intervals of type I error rate of different methods: Control MCV1 rate: 0.7; Coefficient set: 3; $ICC_V$: 0.24*

| Villages per arm | Quasi-binomial | Beta | Naive |
| --- | --- | --- | --- |
| 60 | 0.084 (0.079, 0.090) | 0.043 (0.039, 0.047) | 0.047 (0.043, 0.051) |
| 70 | 0.087 (0.081, 0.092) | 0.047 (0.042, 0.051) | 0.051 (0.047, 0.056) |
| 80 | 0.090 (0.084, 0.095) | 0.046 (0.042, 0.050) | 0.052 (0.048, 0.057) |
| 90 | 0.089 (0.083, 0.095) | 0.051 (0.047, 0.055) | 0.044 (0.040, 0.048) |

*Supplementary Table S26.* *Estimates and 95% confidence intervals of type I error rate of different methods: Control MCV1 rate: 0.7; Coefficient set: 3; $ICC_V$: 1/3*

| Villages per arm | Quasi-binomial | Beta | Naive |
| --- | --- | --- | --- |
| 60 | 0.086 (0.081, 0.092) | 0.041 (0.037, 0.045) | 0.048 (0.044, 0.052) |
| 70 | 0.089 (0.084, 0.095) | 0.040 (0.036, 0.043) | 0.048 (0.044, 0.052) |
| 80 | 0.089 (0.083, 0.094) | 0.040 (0.036, 0.044) | 0.047 (0.043, 0.051) |
| 90 | 0.086 (0.081, 0.092) | 0.040 (0.037, 0.044) | 0.043 (0.039, 0.047) |

*Supplementary Table S27.       Estimates and 95% confidence intervals of type I error rate of different methods: Control MCV1 rate: 0.75; Coefficient set: 1; ICC$_V$: 0.24*

| Villages per arm | Quasi-binomial | Beta | Naive |
|---|---|---|---|
| 60 | 0.089 (0.083, 0.094) | 0.049 (0.044, 0.053) | 0.046 (0.042, 0.050) |
| 70 | 0.090 (0.084, 0.095) | 0.044 (0.040, 0.048) | 0.044 (0.040, 0.048) |
| 80 | 0.085 (0.079, 0.090) | 0.045 (0.041, 0.049) | 0.044 (0.040, 0.049) |
| 90 | 0.085 (0.079, 0.090) | 0.048 (0.044, 0.052) | 0.042 (0.038, 0.046) |

*Supplementary Table S28.       Estimates and 95% confidence intervals of type I error rate of different methods: Control MCV1 rate: 0.75; Coefficient set: 1; ICCV: 1/3*

| Villages per arm | Quasi-binomial | Beta | Naive |
|---|---|---|---|
| 60 | 0.086 (0.080, 0.091) | 0.040 (0.037, 0.044) | 0.043 (0.039, 0.047) |
| 70 | 0.088 (0.082, 0.093) | 0.039 (0.035, 0.043) | 0.046 (0.042, 0.050) |
| 80 | 0.095 (0.089, 0.101) | 0.039 (0.035, 0.042) | 0.046 (0.042, 0.050) |
| 90 | 0.084 (0.079, 0.089) | 0.039 (0.035, 0.043) | 0.043 (0.039, 0.047) |

*Supplementary Table S29.       Estimates and 95% confidence intervals of type I error rate of different methods: Control MCV1 rate: 0.75; Coefficient set: 2; ICC$_V$: 0.24*

| Villages per arm | Quasi-binomial | Beta | Naive |
|---|---|---|---|
| 60 | 0.087 (0.082, 0.093) | 0.044 (0.040, 0.048) | 0.042 (0.039, 0.046) |
| 70 | 0.088 (0.082, 0.094) | 0.038 (0.035, 0.042) | 0.041 (0.037, 0.045) |
| 80 | 0.089 (0.084, 0.095) | 0.036 (0.032, 0.039) | 0.038 (0.035, 0.042) |
| 90 | 0.091 (0.085, 0.097) | 0.042 (0.038, 0.046) | 0.041 (0.037, 0.045) |

*Supplementary Table S30.     Estimates and 95% confidence intervals of type I error rate of different methods: Control MCV1 rate: 0.75; Coefficient set: 2; ICC$_v$: 1/3*

| Villages per arm | Quasi-binomial | Beta | Naive |
|---|---|---|---|
| 60 | 0.088 (0.082, 0.094) | 0.047 (0.043, 0.051) | 0.046 (0.042, 0.051) |
| 70 | 0.089 (0.083, 0.094) | 0.048 (0.044, 0.052) | 0.046 (0.042, 0.050) |
| 80 | 0.093 (0.087, 0.099) | 0.046 (0.042, 0.050) | 0.043 (0.039, 0.047) |
| 90 | 0.086 (0.081, 0.092) | 0.047 (0.043, 0.051) | 0.40    .036, 0.043) |

*Supplementary Table S31.     Estimates and 95% confidence intervals of type I error rate of different methods: Control MCV1 rate: 0.75; Coefficient set: 3; ICC$_v$: 0.24*

| Villages per arm | Quasi-binomial | Beta | Naive |
|---|---|---|---|
| 60 | 0.091 (0.085, 0.096) | 0.048 (0.044, 0.053) | 0.048 (0.043, 0.052) |
| 70 | 0.087 (0.082, 0.093) | 0.048 (0.044, 0.052) | 0.045 (0.041, 0.049) |
| 80 | 0.087 (0.082, 0.093) | 0.044 (0.040, 0.048) | 0.043 (0.040, 0.047) |
| 90 | 0.088 (0.083, 0.094) | 0.046 (0.042, 0.050) | 0.044 (0.040, 0.048) |

*Supplementary Table S32.  Estimates and 95% confidence intervals of type I error rate of different methods: Control MCV1 rate: 0.75; Coefficient set: 3; $ICC_V$: 1/3*

| Villages per arm | Quasi-binomial | Beta | Naive |
| --- | --- | --- | --- |
| **60** | 0.086 (0.081, 0.092) | 0.041 (0.037, 0.045) | 0.042 (0.038, 0.046) |
| **70** | 0.085 (0.080, 0.091) | 0.037 (0.034, 0.041) | 0.043 (0.039, 0.047) |
| **80** | 0.099 (0.093, 0.105) | 0.044 (0.040, 0.048) | 0.049 (0.045, 0.054) |
| **90** | 0.088 (0.083, 0.094) | 0.039 (0.035, 0.042) | 0.041 (0.037, 0.045) |

# 5. Supplementary for Simulations

*Supplementary Figure S5.*     *Power for base case event rate of 0.70 (Beta regression)*

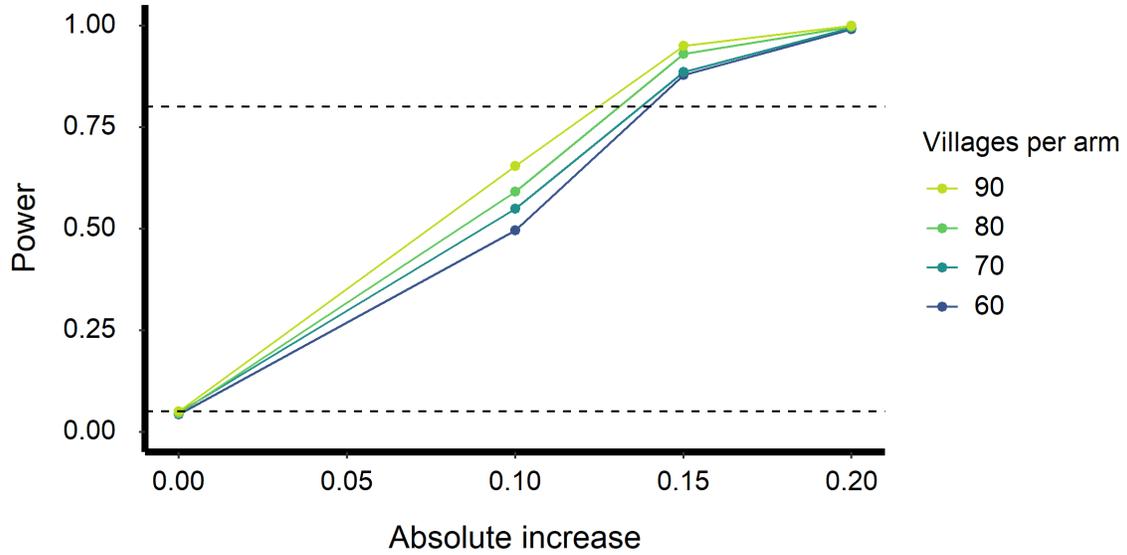

*Supplementary Figure S6.*     *Power for base case event rate of 0.70 (Naïve Analysis)*

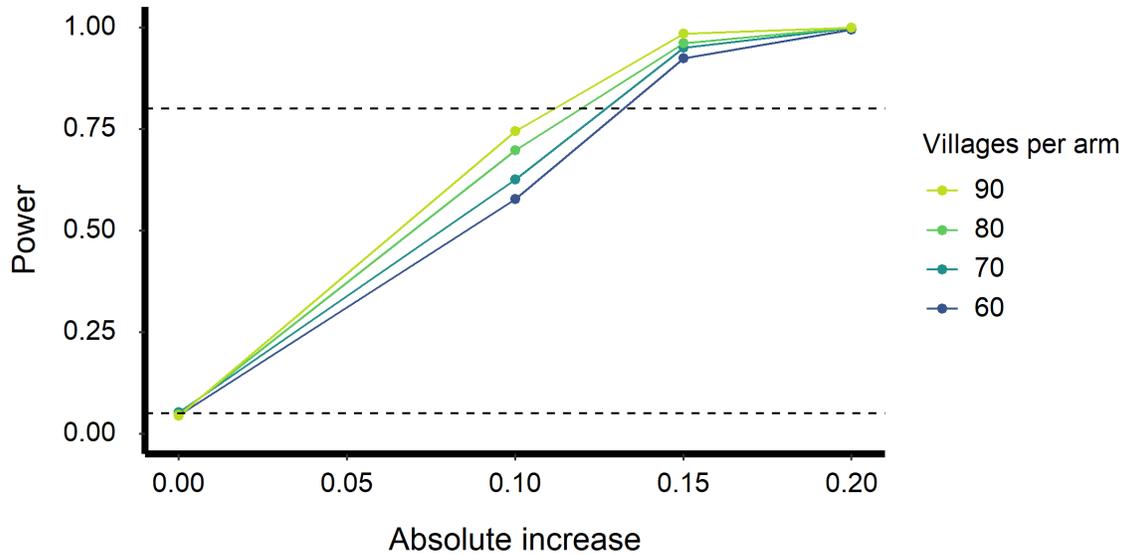

*Supplementary Figure S7.*       *Power of beta regression at different sample sizes and absolute increases in MCV1 rate under different MCV1 rates in the control arm (columns), coefficient sets (rows) and an $ICC_V$ of 0.24*

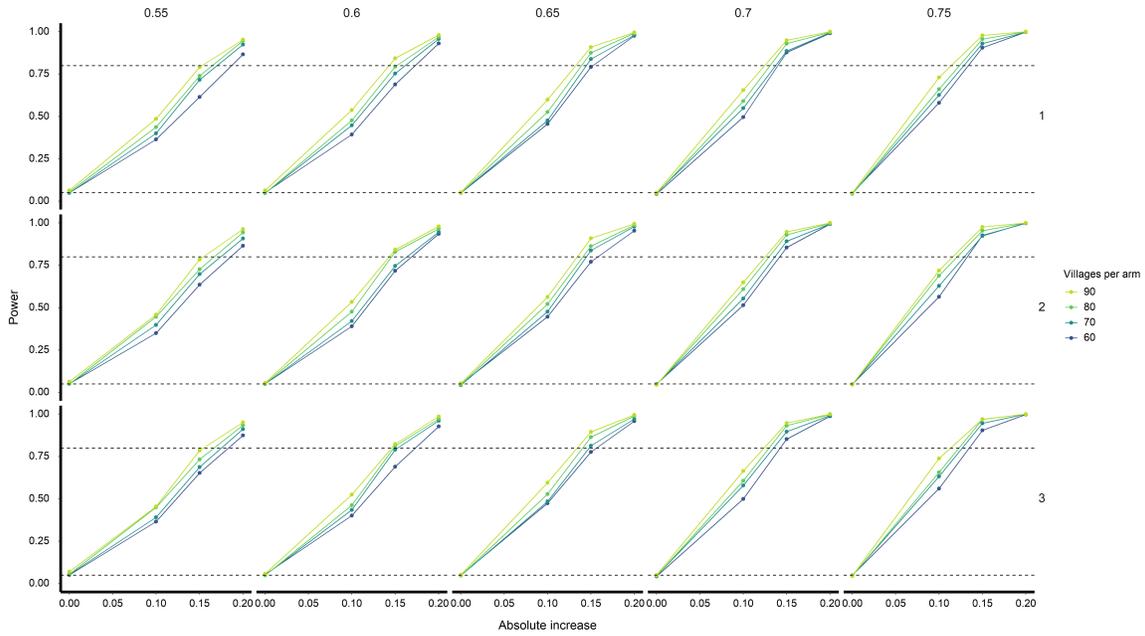

*Supplementary Figure S8.*       *Power of beta regression at different sample sizes and absolute increases in MCV1 rate under different MCV1 rates in the control arm (columns), coefficient sets (rows) and an $ICC_V$ of 1/3*

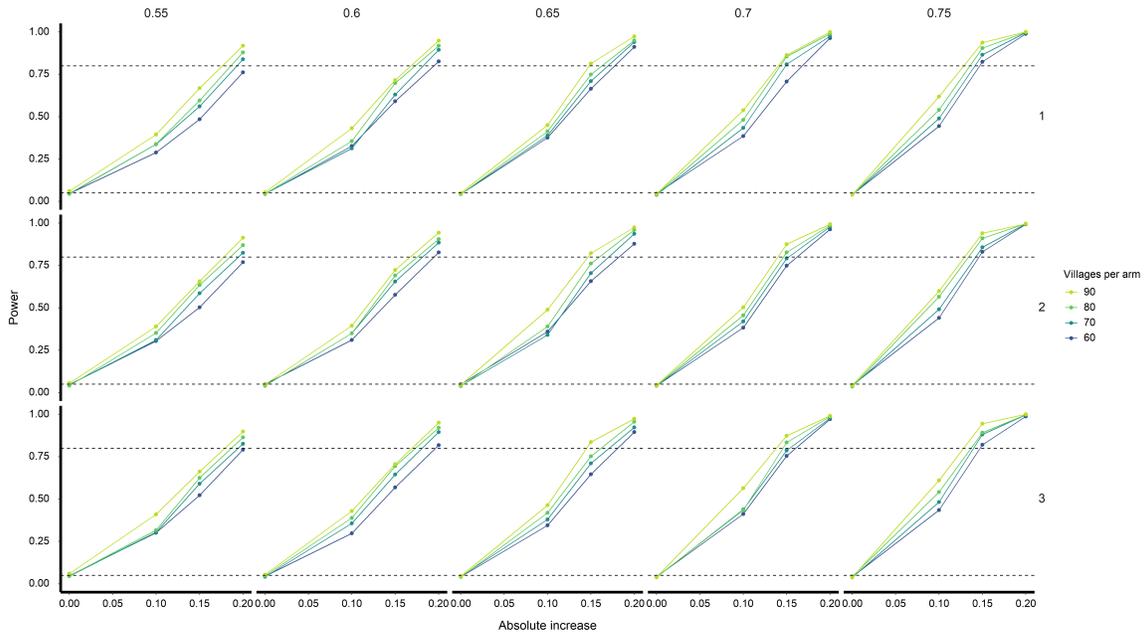

*Supplementary Figure S9.* *Power of quasi-binomial regression at different sample sizes and absolute increases in MCV1 rate under different MCV1 rates in the control arm (columns), coefficient sets (rows) and an $ICC_V$ of 0.24*

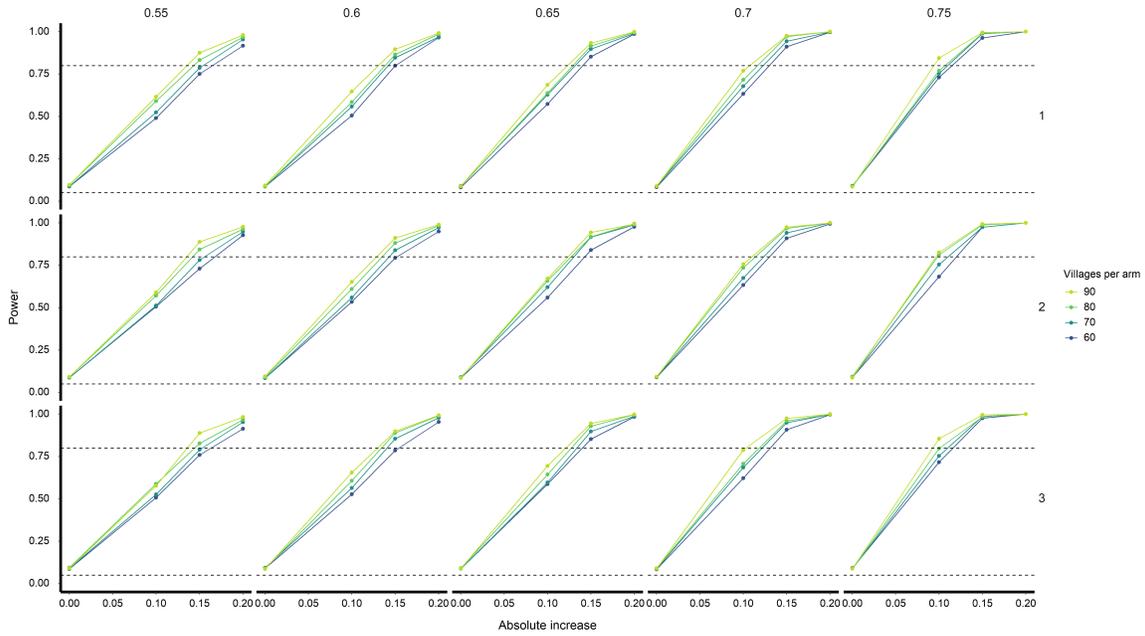

*Supplementary Figure S10.* *Power of quasi-binomial regression at different sample sizes and absolute increases in MCV1 rate under different MCV1 rates in the control arm (columns), coefficient sets (rows) and an $ICC_V$ of 1/3*

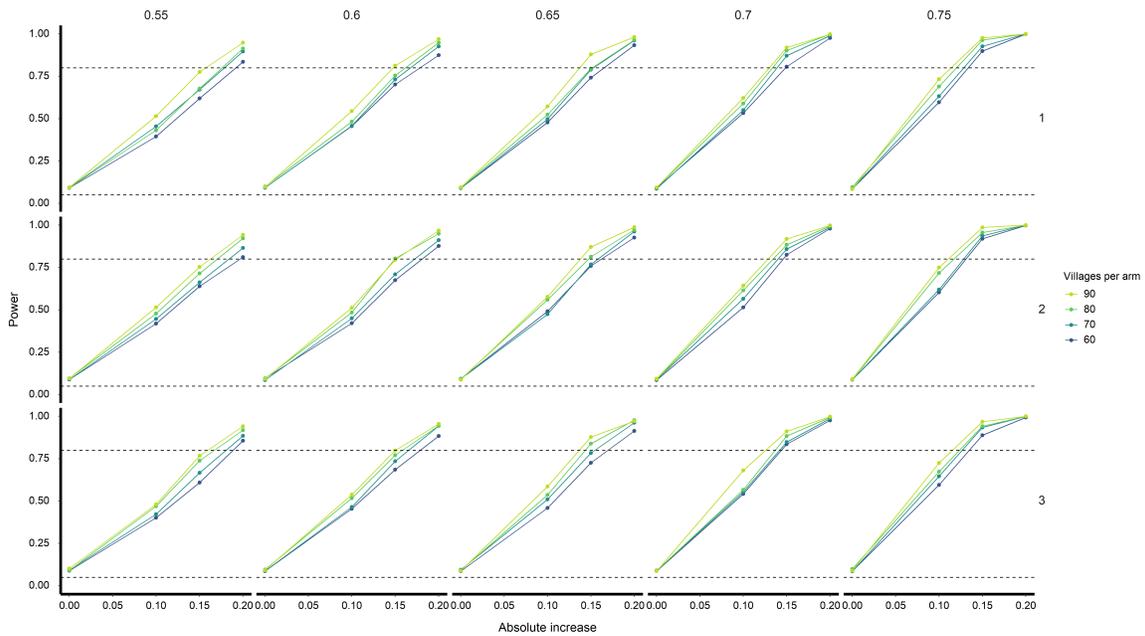

*Supplementary Figure S11.     Power of Naïve analysis at different sample sizes and absolute increases in MCV1 rate under different MCV1 rates in the control arm (columns), coefficient sets (rows) and an $ICC_V$ of 0.24*

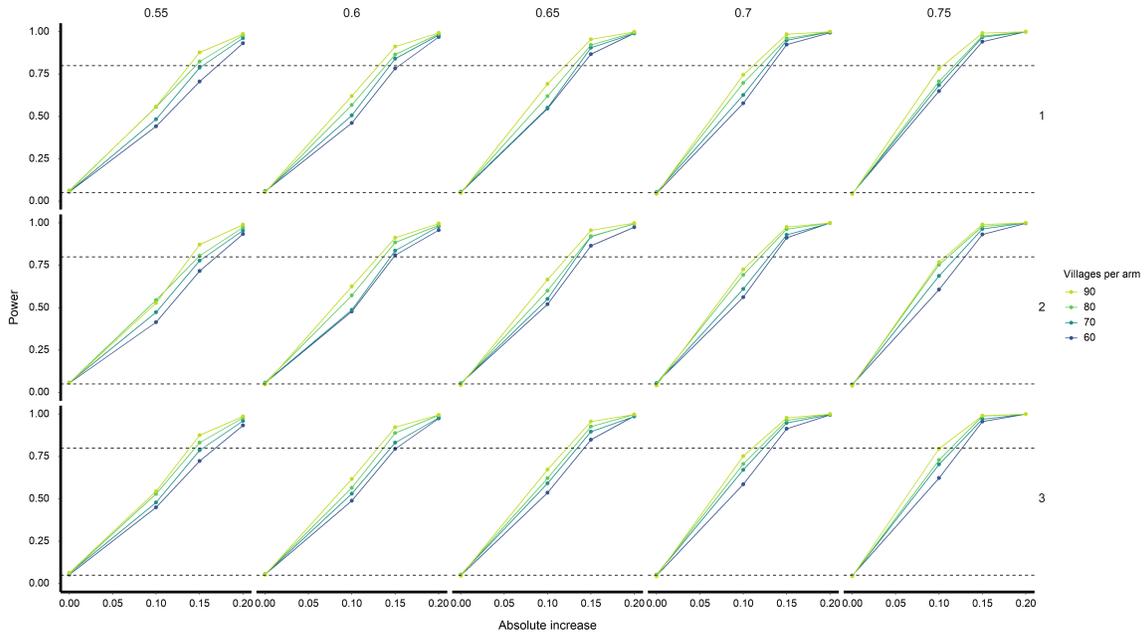

*Supplementary Figure S12.     Power of Naïve analysis at different sample sizes and absolute increases in MCV1 rate under different MCV1 rates in the control arm (columns), coefficient sets (rows) and an $ICC_V$ of 1/3*

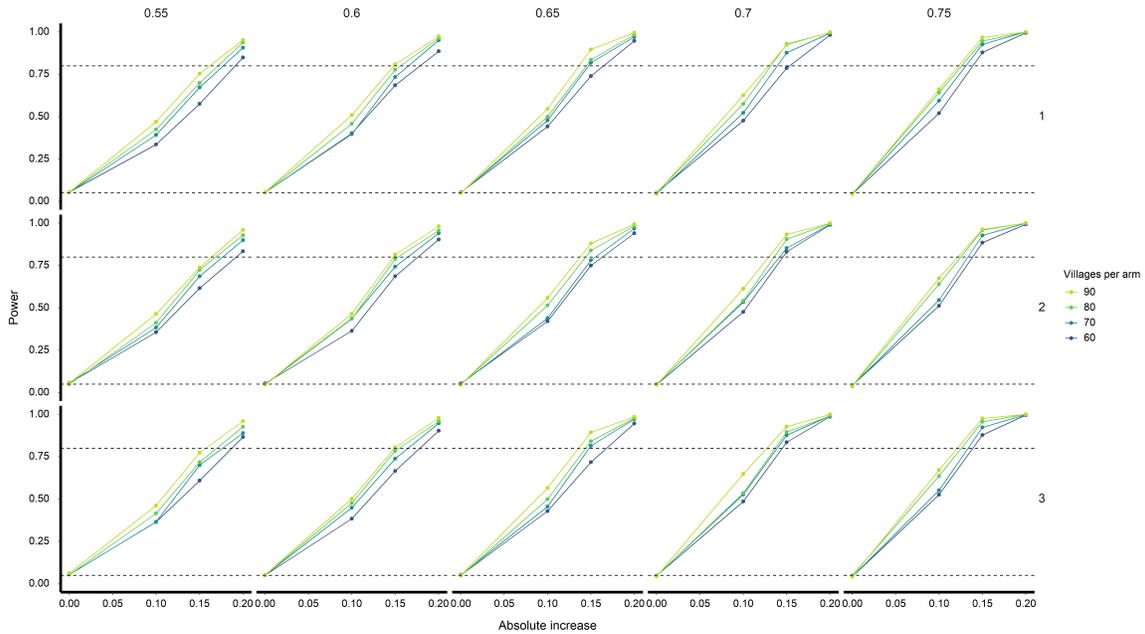